\documentclass{aa}

\usepackage{txfonts}

\usepackage{graphicx}

\usepackage{epsfig}

\usepackage{natbib}
\bibpunct{(}{)}{;}{a}{}{,} 

\newcommand{\teff}{$T_\mathrm{eff}$}

\newcommand{\rwd}{$R_\mathrm{WD}$}

\newcommand{\chisq}{$\chi^2$}
\newcommand{\chisqred}{$\chi^2_\nu$}

\newcommand{\halpha}{\mbox{H$\alpha$}}
\newcommand{\hbeta}{\mbox{H$\beta$}}

\newcommand{\glm}{$g_l^m$}
\newcommand{\hlm}{$h_l^m$}

\newcommand{\oef}{\mbox{\object{EF~Eri}}}
\newcommand{\obl}{\mbox{\object{BL~Hyi}}}
\newcommand{\ocp}{\mbox{\object{CP~Tuc}}}

\newcommand{\eferi}{\mbox{\object{EF~Eri}}}
\newcommand{\blhyi}{\mbox{\object{BL~Hyi}}}
\newcommand{\cptuc}{\mbox{\object{CP~Tuc}}}

\newcommand{\bpsi}{$B-\psi$}

\begin{document}

\title{Zeeman tomography of magnetic white dwarfs \subtitle{IV. The
complex field structure of the polars
EF~Eri, BL~Hyi, and CP~Tuc\thanks{Based on observations
collected at the European Southern Observatory, Paranal, Chile, under
programme IDs \mbox{63.P-0003(A)}, \mbox{64.P-0150(C)}, and \mbox{66.D-0128(B)}.}}}

\author{ K.~Beuermann\inst{1}, F.~Euchner\inst{1}\fnmsep\thanks{Present
address: Swiss Seismological Service, ETH H\"onggerberg, CH-8093
Z\"urich, Switzerland}, K.~Reinsch\inst{1}, S.~Jordan\inst{2}, \and
B.\,T.~G\"ansicke\inst{3}}


  
 \institute{Institut f\"ur Astrophysik, Universit\"at G\"ottingen,
 Friedrich-Hund-Platz~1, D-37077~G\"ottingen, Germany, e-mail:
 beuermann@astro.physik.uni-goettingen.de \and Astronomisches
 Rechen-Institut am ZAH, \mbox{M\"onchhofstr.~12--14},
 \mbox{D-69120~Heidelberg}, Germany, e-mail:
 jordan@ari.uni-heidelberg.de \and Department of Physics, University
 of Warwick, Coventry~CV4~7AL, UK, e-mail: Boris.Gaensicke@warwick.ac.uk}


\date{Received September 1, 2006\ / Accepted September 29, 2006}
  
\abstract { 
The magnetic fields of the accreting white dwarfs in magnetic
cataclysmic variables (mCVs) determine the accretion geometries, the
emission properties, and the secular evolution of these objects.}
{ 
We determine the structure of the surface magnetic fields of the
white dwarf primaries in magnetic CVs using Zeeman tomography.}
{ 
Our study is based on orbital-phase resolved optical flux and circular
polarization spectra of the polars \oef, \obl, and \ocp\ obtained with
FORS1 at the ESO VLT. An evolutionary algorithm is used to synthesize
best fits to these spectra from an extensive database of pre-computed
Zeeman spectra. The general approach has been described in previous
papers of this series.}
{ 
The results achieved with simple geometries as centered or offset
dipoles are not satisfactory. Significantly improved fits are obtained
for multipole expansions that are truncated at degree
$l_\mathrm{max}=3$ or 5 and include all tesseral and sectoral
components with $0\le m\le l$.  The most frequent field strengths of
13, 18, and 10\,MG for \oef, \obl, and \ocp\ and the ranges of field
strength covered are similar for the dipole and multipole models, but
only the latter provide access to accreting matter at the right
locations on the white dwarf. The results suggest that the field
geometries of the white dwarfs in short-period mCVs are quite complex
with strong contributions from multipoles higher than the dipole in
spite of a typical age of the white dwarfs in CVs in excess of 1~Gyr. }
{ 
It is feasible to derive the surface field structure of an accreting
white dwarf from phase-resolved low-state circular spectropolarimetry
of sufficiently high signal-to-noise ratio. The fact that independent
information is available on the strength and direction of the field in
the accretion spot from high-state observations helps in unraveling
the global field structure.  }

\titlerunning{Magnetic field structure of white dwarfs in CVs}
\authorrunning{K.~Beuermann et al. }

\maketitle


\section{Introduction}

The subclass of magnetic cataclysmic variables (mCVs) termed polars
\citep{krzeminskiserkowski77} contains an accreting white dwarf that
emits circularly polarized cyclotron radiation from an accretion
region standing off the photosphere, often referred to as accretion
spot. The harmonic structure of the cyclotron radiation allows a
straightforward measurement of the magnetic field and an estimate of
the field direction in the spot. Photospheric absorption lines are
heavily veiled by the intense cyclotron emission in the high
(accreting) state. The field structure over the surface of the white
dwarf becomes accessible to measurement only in low states of
discontinued accretion via the profiles of the photospheric
Zeeman-broadened Balmer absorption lines, an approach which is
applicable also to non-accreting isolated white dwarfs. Different from
the latter, accreting systems offer the advantage that the strength
and direction of the field in the accretion spot and its approximate
location on the surface as determined from high-state observations
represent a fix point for the field structure.
For simplicity it was often assumed that the field is quasi-dipolar,
although accreting systems with two accretion spots separated by much
less than $180^\circ$ supported suspicions of a more complex structure
(Meggitt \& Wickramasinghe 1989, Piirola et al 1987b, see
Wickramasinghe \& Ferrario 2000 for a review).

We have set out on a program to obtain a more complete picture of the
surface field structure of magnetic white dwarfs using an approach
dubbed Zeeman tomography \citep{euchneretal02}.  The field geometries
of two isolated white dwarfs, HE\,1045-0908 and PG1015+014
\citep{euchneretal05,euchneretal06}, proved to be significantly more
complex than simple centered or offset dipoles. In this paper, we
present first results of the Zeeman tomography of three polars
observed in their low states, \oef, \obl, and \ocp, and find that
they, too, have rather complex field geometries.


\section{Observations and Data Analysis}

We have obtained spin phase-resolved circular spectropolarimetry of
\oef, \obl, and \ocp\ in their low states. These stars belong to the
short-period variety with orbital periods of 81.0\,min (\eferi),
113.6\,min (\blhyi), and 89.0\,min (\cptuc). The secondary star is a
late M-star in \blhyi. It is substellar in \eferi\
\citep{beuermannetal00, harrisonetal04} and possibly in \cptuc,
too. Full orbital coverage was achieved for \eferi\ and \blhyi, but
for technical reasons only half of the orbital period was covered for
\cptuc.

The data were collected at the ESO VLT using the focal reducer
spectrograph FORS1. The instrument was operated in spectropolarimetric
(PMOS) mode, with the GRIS\_300V+10 grism and an order separation
filter GG~375, yielding a usable wavelength range of 3750--8450\,\AA.
With a slit width of 1\arcsec\, the FWHM spectral resolution was
\mbox{13\,\AA}\ at \mbox{5500\,\AA}. A signal-to-noise ratio of
typically $S/N \simeq 100$ was reached for the individual flux
spectra. FORS1 contains a Wollaston prism for beam separation and two
superachromatic phase retarder plate mosaics. Since both plates cannot
be used simultaneously, only the circular polarization has been
recorded using the quarter wave plate. Spectra of the target star and
comparison stars in the field have been obtained simultaneously by
using the multi-object spectroscopy mode of FORS1. This allows us to
derive individual correction functions for the atmospheric absorption
losses in the target spectra and to check for remnant instrumental
polarization. Tab.~1 contains a log of the observations.



\begin{table}[b]
\caption{Dates of the spectropolarimetric observations obtained at the
ESO VLT, exposure times, and number of exposures. }
\label{tab:obs-log}
\centering
\begin{tabular}{l@{\hspace{6mm}}ll@{\hspace{6mm}}l@{\hspace{6mm}}c}
\hline \hline \noalign{\smallskip}
Object & Date & UT & $t_\mathrm{exp}$ (s) & number \\
\noalign{\smallskip} \hline
\noalign{\smallskip}
\oef	& 2000/11/22 & 01:02--03:03 & 360 & 14 \\
    	&            & 04:24--05:14 & 360 & \hspace{1.4mm}6  \\
\obl	& 1999/12/04 & 04:25--06:29 & 360 & 16 \\
\ocp	& 1999/06/04 & 08:07--08:44 & 480 & \hspace{1.4mm}4  \\
    	&            & 09:35--10:31 & 480 & \hspace{1.4mm}6  \\
\noalign{\smallskip} \hline
\end{tabular}
\end{table}

The observational data have been reduced according to standard
procedures (bias, flat field, night sky subtraction, wavelength
calibration, atmospheric extinction, flux calibration) using the
context MOS of the ESO MIDAS package.
In order to eliminate observational biases caused by Stokes parameter
crosstalk, the wavelength-dependent degree of circular polarization
$V/I$ has been computed from two consecutive exposures recorded with
the quarter wave retarder plate rotated by $\pm$45\degr. The circular
polarization was then obtained as the average of two consecutive sets
of spectra in the ordinary and the extraordinary beams \citep[see][for
details]{euchneretal05}. 

\begin{figure}[t]
\includegraphics[width=8.8cm]{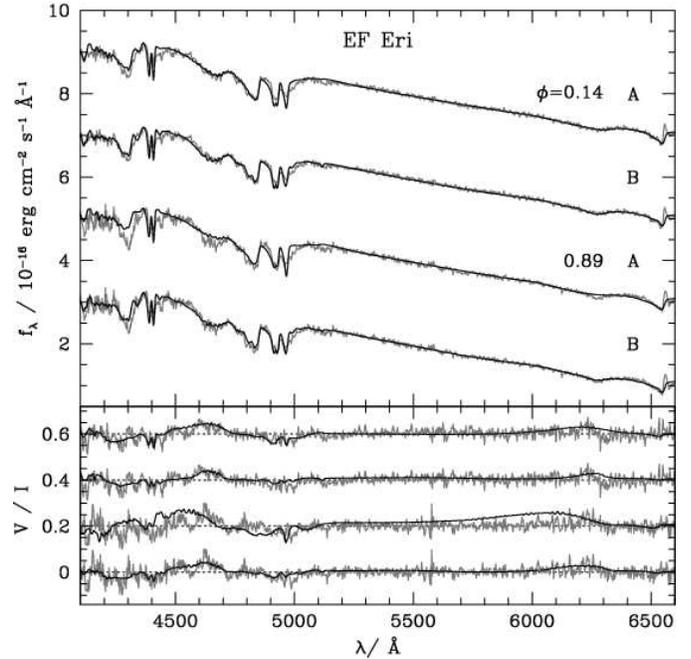}
\caption{Flux spectra (top) and circular polarization spectra (bottom)
of \eferi\ at two selected orbital phases and for two field
models. Model A is an offset dipole and model B a multipole expansion
truncated at $l_\mathrm{max}=5$. The data are shown as grey curves,
the best fit models from the Zeeman tomographic analysis are overlaid
as solid black curves. The order of orbital phases and field models
for the polarization spectra is the same as for the flux spectra,
i.e., the bottom spectrum and third from the bottom are for the
multipole expansion, the other two for the offset dipole. The ordinate
scales refer to the bottom spectrum, the other ones are arbitrarily
shifted upwards.}
\label{fig:spec1}
\end{figure}

The three stars were in their low states with magnitudes estimated
from the spectrophotometry of $V\sim 18$ for \eferi\ and \blhyi\ and
$V\sim 19$ for \cptuc. CCD photometry of \blhyi\ in the same night
gave $V=17.45$. 
In case of \blhyi, the resulting spectra were corrected for the
contribution by the secondary star using a spectrum of the dM5.5 star
Gl\,473 as a template. No trace of the secondary star was seen in the
other two objects. Seeing variations and a loss of blue flux in the first two hours of
the \eferi\ run and in the last three spectra of \blhyi\
affected the detection of orbital modulations. The
orbital modulation of \eferi\ seen in the remainder of the data is
consistent with that reported by \citet[][and references
therein]{szkodyetal06}. No substantial orbital modulation was detected
in the data of \blhyi\ and \cptuc.

\begin{figure*}[t]
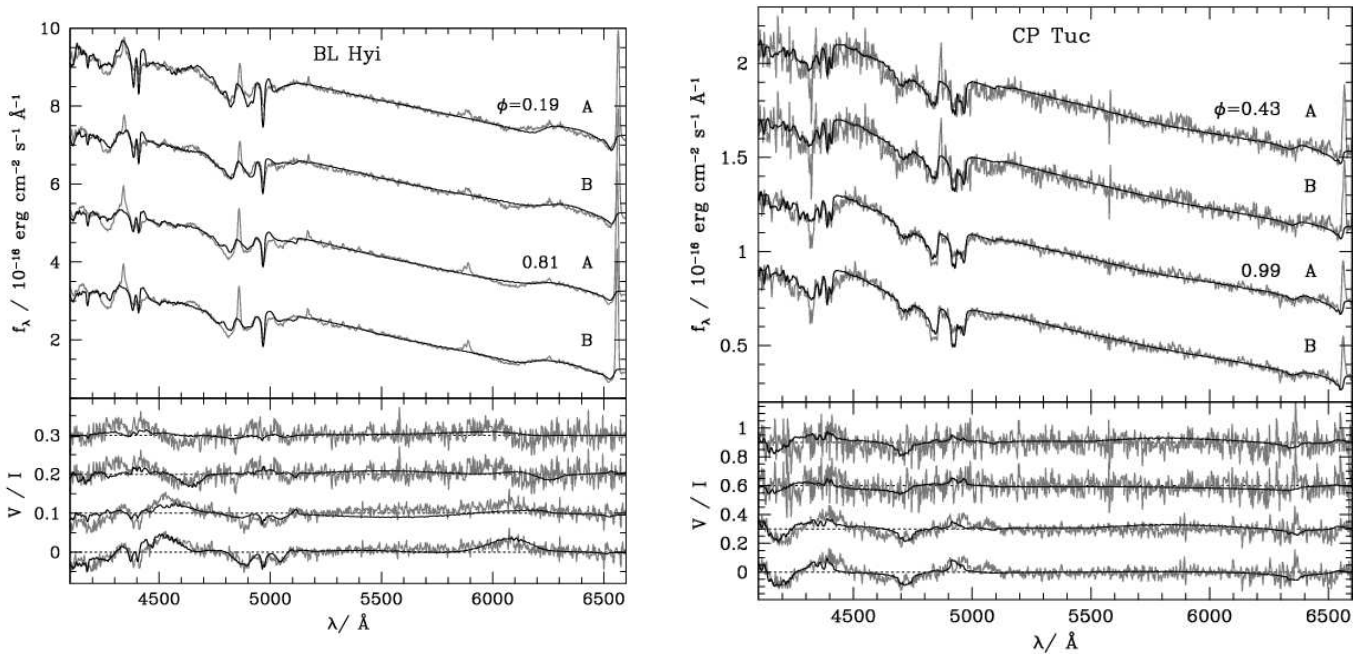

\includegraphics[width=8.8cm]{6332.f2a}
\hfill
\includegraphics[width=8.8cm]{6332.f2b}
\caption{As Fig. 1, but for \blhyi\ and \cptuc. Model A is an offset
dipole and model B a multipole expansion truncated at $l_\mathrm{max}=3$.}
\label{fig:spec2}
\end{figure*}

In order to facilitate analysis of the Zeeman absorption features, the
continua of the observed spectra were normalized by the following
procedure which minimizes the differences between observed and
theoretical continua and corrects also for the mentioned loss of blue
light. In a first step, a mean effective temperature of each object
was determined by fitting a magnetic model spectrum to the
quasi-continuum of the mean of the observed spectra unaffected by
light loss. In a second step, the continua of all observed flux
spectra were sampled in some 20 narrow fiducial wavelength intervals,
which avoid the known Zeeman features and the emission lines, and
adjusted to the best-fitting model spectrum using low-order
polynomials \citep[see][for more details]{euchneretal05}. This
approach removes the time variability in the flux continua but leaves
the equivalent widths of the Zeeman features largely unaffected.  The
mean effective temperature is \teff~=~11000\,K for \eferi, 12000\,K
for \blhyi, and 10000\,K for \cptuc, with an estimated accuracy of
about \mbox{$\sim1500$\,K}. Using this mean temperature in the
tomographic analysis affects the theoretical Zeeman absorption
features only minimally, because their equivalent width reaches a
maximum around 11000\,K and varies \mbox{little} with effective temperature
around the maximum. In passing we note that our mean effective
temperatures confirm the rather low temperatures of the white dwarfs
in polars \citep[][and references therein]{araujobetancoretal05}.

For the tomographic analysis, the flux and polarization spectra were
collected into $n=4$ phase bins for \eferi\ and $n=5$ for \blhyi\
about equally spaced to cover the whole orbit. The spectra from the half
orbit of \cptuc\ were gathered into three bins.

\section{General approach}

We determine the global surface magnetic field structure using the
Zeeman tomographic procedure described by
\citet{euchneretal02,euchneretal05,euchneretal06}.  The process
involves the inversion of the one-dimensional time series of
rotational phase-resolved Zeeman flux and circular polarization
spectra obtained in a non-accreting (low) state into a two-dimensional
field distribution over the surface of the star. Because of the finite
signal-to-noise ratio, this inversion problem may allow more than one
solution within the observational uncertainties \citep[for a
discussion see][]{euchneretal02}.
This ambiguity arises from the fact that different models
may have similar frequency distributions of the field strength $B$ and
the angle $\psi$ between field vector and line of sight and, hence,
yield similar Zeeman spectra, but differ in the arrangement of the
field over the surface. In this situation, it is advantageous to use
the field vector in the accretion spot as deduced from cyclotron
spectroscopy and broad-band circular polarimetry in a high state as a
fixpoint and effective constraint on the tomographic procedure. In the
present paper, we have not included such a constraint in a formal way,
but use it to select between solutions of the tomographic process
obtained for different assumptions on the field geometry. To this end,
we follow the surface field outward and determine the maximal radial
distance reached by each field line. Field lines extending to more
than $R_\mathrm{max}=10$\,\rwd\ are considered 'open'. The accretion
stream can couple to field lines which reach out sufficiently far,
with the actual radius at which coupling can occur depending on the
ram pressure of the accreting matter and the local field strength. The
requirement that field lines which originate at a specific point at
the surface reach out to more than several white dwarf radii can
effectively discriminate between different field models. To be sure, a
model which provides a good fit to the Zeeman spectra \emph{and}
possesses field lines reaching far out at the required position may
not be \emph{the correct model}, but is as close to reality as we can
presently get. As a further caveat note that the actual coupling
conditions have not been investigated and part of the far-reaching field
lines may not be accessible to the stream.  Consequently, only a
fraction of the long ribbon-like structures of far-reaching field lines
which appear in some models may act as foot points of accreting field
lines. Nevertheless, with this information included, the analysis of
accreting white dwarfs may yield more definite results than that of
isolated white dwarfs.

\begin{table*}[t]
\caption{Best-fit magnetic parameters for the truncated multipole
expansions up to degree \mbox{$l_\mathrm{max}=5$} for \eferi\ and up
to \mbox{$l_\mathrm{max}=3$} for \blhyi\ and \cptuc. The coefficients
\glm\ and \hlm\ are in MG. The tilt angle of the multipole axis
relative to the rotational axis is \mbox{74\degr}, \mbox{32\degr},
and \mbox{23\degr} for \eferi, \blhyi, and \cptuc, respectively.}
\label{tab:multipole}
\centering
\begin{tabular}{@{\hspace{5mm}}cc@{\hspace{10mm}}rrrrr@{\hspace{10mm}}rrr@{\hspace{10mm}}rrrr} 
\hline \hline \noalign{\smallskip}
   &      & \multicolumn{5}{c}{\eferi}  &  \multicolumn{3}{c}{\blhyi}  &  \multicolumn{3}{c}{\cptuc} \\
 $m$ &    & $l=1$   & 2       & 3       & 4       & 5       &  $l=1$   & 2       & 3       & $l=1$   & 2       & 3       \\ 
             \noalign{\smallskip} \hline \noalign{\smallskip}
 0 & $g_l^0$ & $-$5.9  & 4.2     & 4.1     & $-$1.9  &$-$0.3   & $-$5.2   & $-$12.0 & 5.8     & 3.5     & 14.6    & $-$1.6  \\
 1 & $g_l^1$ & 6.1     & $-$4.8  & 1.2     & 1.0 	  &$-$0.4   & $-$8.1   & 15.8    & $-$2.5  & $-$15.5 & 8.5     & $-$2.8  \\
   & $h_l^1$ & 0.7     & 12.7    & $-$1.2  & 0.2  	  &4.9      & 0.2      & 12.8    & 6.6     & 8.1     & $-$13.0 & 4.9     \\
 2 & $g_l^2$ &         & 10.8    & $-$3.9  & $-$4.0  &$-$1.6   &          & 12.5    & $-$2.7  &         & $-$3.7  & $-$1.5  \\
   & $h_l^2$ &         & $-$1.0  & 5.4     & $-$5.8  &2.1      &          & $-$10.4 & $-$2.1  &         & 0.3     & $-$1.7  \\
 3 & $g_l^3$ &         &         & 7.7     & $-$2.0  &$-$0.9   &          &         & $-$2.5  &         &         & $-$4.9  \\
   & $h_l^3$ &         &         & 1.8     & $-$5.1  &$-$0.6   &          &         & 8.6     &         &         & 0.0     \\
 4 & $g_l^4$ &         &         &         & 2.4     &$-$0.2   &          &	 & 	   &         &         &         \\
   & $h_l^4$ &         &         &         & 1.5     &3.7      &	       &         &	   &	     &	       &         \\
 5 & $g_l^5$ &         &         &         &         &$-$1.2   &  	       &	 &	   &	     &	       &         \\
   & $h_l^5$ &         &         &         &         &3.2      &  	       &         &	   &	     &	       &	 \\ [0.7ex]
\noalign{\smallskip} \hline
\end{tabular}
\end{table*}

\begin{figure*}[t]
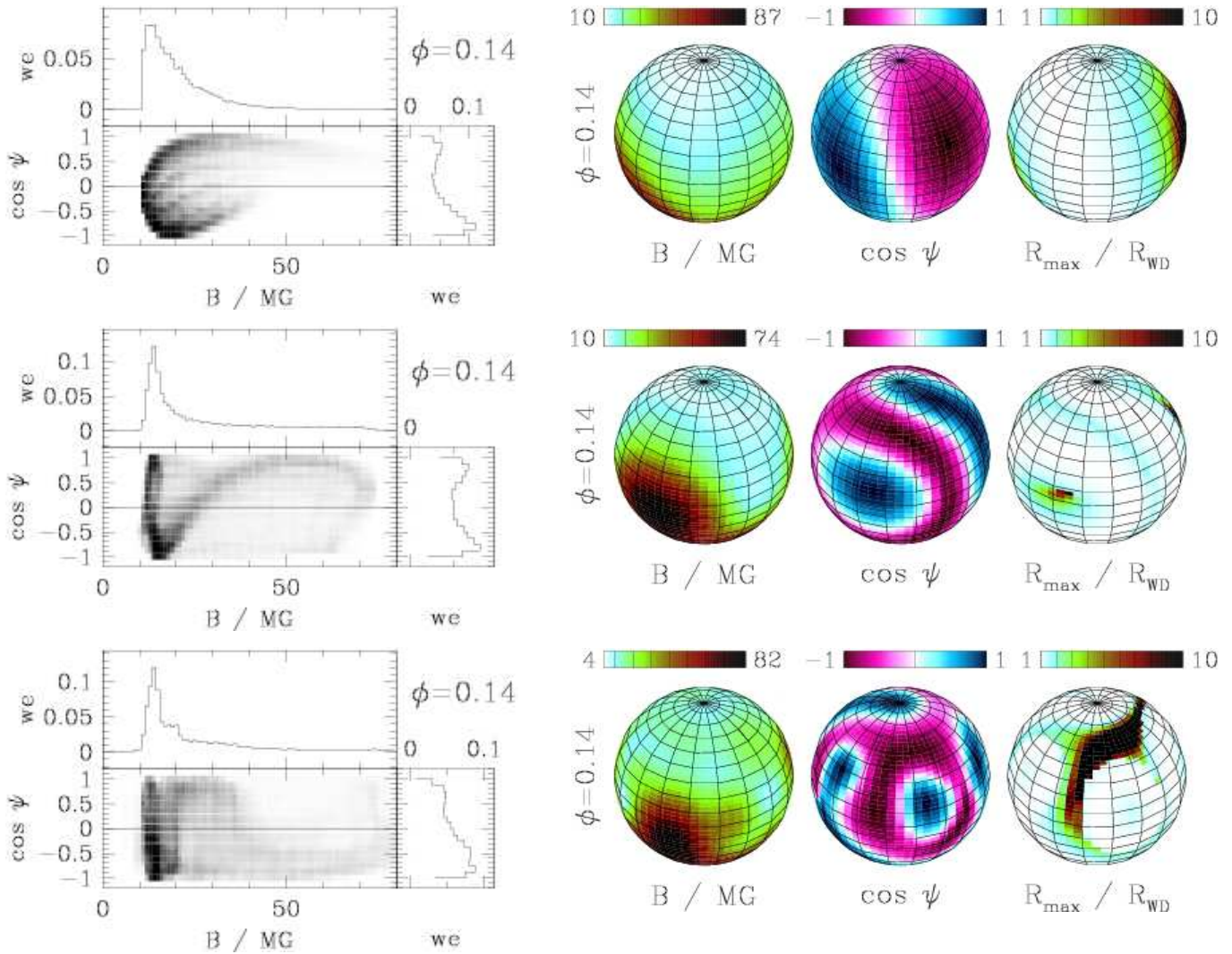

\includegraphics[width=7.6cm,clip]{6332.f3a}
\hfill
\raisebox{6mm}{\includegraphics[width=9.6cm,clip]{6332.f3b}}\\[1.5ex]
\includegraphics[width=7.6cm,clip]{6332.f3c}
\hfill
\raisebox{6mm}{\includegraphics[width=9.6cm,clip]{6332.f3d}}\\[1.5ex]
\includegraphics[width=7.6cm,clip]{6332.f3e}
\hfill
\raisebox{6mm}{\includegraphics[width=9.6cm,clip]{6332.f3f}}
\caption{Visualization of the three magnetic field models discussed
for \eferi. The results shown are for orbital phase $\phi=0.14$ near
the peaks of the X-ray and infrared cyclotron fluxes and shortly past
the infrared absorption dip ($\phi=0$). The models are (i) the offset
dipole (top), (ii) the full multipole expansion truncated at
$l_\mathrm{max}=3$ (center), and (iii) the same for $l_\mathrm{max}=5$
(bottom). The left panels show the \bpsi~diagrams in which the
frequency distribution of field vectors over the surface is
represented by a grey scale. Also shown are the weight distributions
(we) projected onto the $B$ and the cos\,$\psi$ axes. The right panels
display globes with the distributions of the absolute value of the
field strength $B$, of cos\,$\psi=B_l\,/B$ with
$B_l$ the field component along the line of sight, and of the
maximum radius in units of \rwd\ to which the field lines
extend. Lines reaching 10\,\rwd\ were not followed further out. The
color bars above the globes indicate the range of the respective
parameters over the visible hemisphere (see the online version of the
paper for colored figures). Progressing orbital phase corresponds to
an anti-clockwise rotation of the globes.}
\label{fig:ef}
\end{figure*}

As in our previous papers on isolated white dwarfs
\citep{euchneretal05, euchneretal06}, we fit the data with either a
hybrid model or a multipole expansion truncated at a maximum degree
$l=l_\mathrm{max}$ including all $l_\mathrm{max}(l_\mathrm{max}+2)$
components with $m=0\ldots l\,$. As a hybrid model, we consider the
superposition of zonal ($m=0$) multipole components which are allowed
to be inclined to each other and to be offset from the center of the
white dwarf. Examples are, e.g., an offset dipole or the sum of dipole
and quadrupole etc.  Such combinations can easily be visualized given
the polar field strengths and orientations of the individual
components. In case of the multipole expansion, on the other hand, the
parameters of the basic dipole are easily interpreted, but the
structure created by the higher multipole components is more difficult
to judge \citep{euchneretal02}. The hybrid models correspond to
special situations not encountered in truncated multipole expansions
of low $l_\mathrm{max}$ and we lack information from dynamo theory on
the feasibility of such models. For the sake of economy and simplicity
of presentation, we present results for the offset dipole as a simple
and popular model and for the full multipole expansion for either
$l_\mathrm{max}=3$ or 5, with occasional comments on other models (a
multipole expansion up to $l_\mathrm{max}=4$ was not tested). We use
two graphic forms to present the results: (i) the '\bpsi~diagrams' which
depict the frequency distribution of field vectors over the surface of
the star at a given orbital phase in the \mbox{$B$--cos\,$\psi$ plane}; and
(ii) actual images of the field distribution. The latter include (a)
the field strength $B$, (b) cos\,$\psi=B_l/B$ with
$B_l$ the field component along the line of sight, and (c) an
image of the maximum radial distance to which a field line extends
that originates from a certain location on the star.

We subjected the flux and polarization spectra at all orbital phases
simultaneously to the tomographic analysis, weighing all wavelengths
equally except for narrow intervals around the Balmer emission
lines. An improved fit can be obtained by restricting it to the set of
flux and polarization spectra at a single phase, but if the model
parameters deduced for different phases disagree there is no unique
solution \citep{euchneretal06}.

\section{Results}

The models are fitted to the average flux and circular polarization
spectra in four orbital phase bins for \eferi, five bins for \blhyi\
and three bins for the half orbit of \cptuc.
The orbital phase conventions used in this paper are the dip ephemeris
for \eferi\ \citep{piirolaetal87a} slightly updated by including the
ROSAT PSPC dip timings from July
1990\,\footnote{$T_0=\mathrm{HJD}244\,3944.9518(6)+0.056265949(14)$\,E,
with 90\% confidence errors.} \citep{beuermannetal91}, the ephemeris
for the start of the bright phase for \blhyi\ \citep{wolffetal99}, and
the dip ephemeris for \cptuc\ \citep{ramsayetal99}.

In our previous papers on the field structure of single white dwarfs,
we considered the inclination of the line of sight relative to the
rotation axis as a free parameter of the fit. For the mCVs, however,
independent and better information on $i$ is available from the light
curve and broad-band polarization studies in their high states. We use
$i=55^\circ$ for \eferi\ \citep{piirolaetal87b,achilleosetal92} and
$i=40^\circ$ for \cptuc\ \citep{thomasreinsch96,ramsayetal99}. For
\blhyi, we use $i=32^\circ$ \citep{schwopeetal95}.

\begin{table}[b]
\caption{Dipole field strength $B_\mathrm{dip}$, most frequent
photospheric field strength $B_\mathrm{prob}$, and ranges
$B_\mathrm{min}$ to $B_\mathrm{max}$ in the models of \oef, \obl, and
\ocp\ in MG. The former refers to the orbital phase when the accretion
spot faces the observer (Fig.~\ref{fig:ef}--\ref{fig:cp}), the latter
to all orbital phases combined.}
\label{tab:brange}
\centering
\begin{tabular}{llccrr}
\hline \hline \noalign{\smallskip}
Object & Model & $B_\mathrm{dip}$ & $B_\mathrm{prob}$ & $B_\mathrm{min}$ 
& $B_\mathrm{max}$ \\
\noalign{\smallskip} \hline
\noalign{\smallskip}
\oef   & offset dipole                & 44.0 & 12 & 10 & 110 \\
       & $l_\mathrm{max}=5$~~multipole &      & 13 & 4  & 111 \\
\obl   & offset dipole                 & 59.5 & 17 & 15 & 88 \\
       & $l_\mathrm{max}=3$~~multipole &      & 18 & 13 & 110 \\
\ocp   & offset dipole                 & 19.8 & 10 &  9 & 41 \\
       & $l_\mathrm{max}=3$~~multipole &      & 10 &  1 & 69 \\
\noalign{\smallskip} \hline
\end{tabular}
\end{table}

Our Zeeman tomography uses the observed flux and circular polarization
spectra of all orbital phases. For conciseness, however, we show the
spectra only for two selected phases and the globes representing the
field structure only for $\phi\simeq \phi_0$ when the main accretion
spot most directly faces the observer, i.e. $\phi_0\simeq 0.10$ for
\eferi, $\phi_0\simeq 0.20$ for \blhyi, and probably $\phi_0\simeq
0.50$ for \cptuc. Our phases closest to $\phi_0$ are $\phi=0.14$ for
\eferi, $\phi=0.19$ for \blhyi, and $\phi=0.43$ for \cptuc.

Figures~\ref{fig:spec1} and \ref{fig:spec2} show the flux and circular
polarization spectra (grey curves) for the value of $\phi$ closest to
$\phi_0$ and for another phase selected to point out differences in
the Zeeman spectra. The spectral data in Figs.~\ref{fig:spec1} and
\ref{fig:spec2} are shown twice along with the best fit theoretical
spectra for two models, the shifted dipole (model A) and the multipole
expansion with $l_\mathrm{max}=3$ or 5 (model B). The ordinate scales
refer to the bottom spectra, the other ones being shifted upwards by
arbitrary amounts. Since the fit to \halpha\ $\sigma^+$ always mimics
that of the $\sigma^-$ feature we have omitted the former in the
figures to avoid excessive compression in wavelength. That feature is
included in the fits, however.

We judged the fits by eye and by a formal global \chisq\ for the flux
and polarization spectra at all phases combined and with all
wavelengths weighted equally. As discussed by \citet{euchneretal05}
the formal reduced \chisqred\ are large, because the adjustment of the
observed continua to the model continua is not perfect and the
standard deviations used in calculating \,\chisq\ account for the
statistical noise in the data but not for the remaining systematic
differences between model and data. We quote the global \chisqred\
values which serve as a guide line, but also judge the merits and
failures of individual models by eye and describe them in words. Not
surprisingly, significantly reduced \chisqred\ are obtained by
excluding the poorly fitting spectral regions from the fit. However,
since these regions differ from object to object, we have refrained
from including such restriction in a general way. We have assured
ourselves, however, that the inclusion of the poorly fitting regions
does not affect the selection of the best-fitting model as the one
with the lowest \chisqred. 

\begin{figure*}[t]
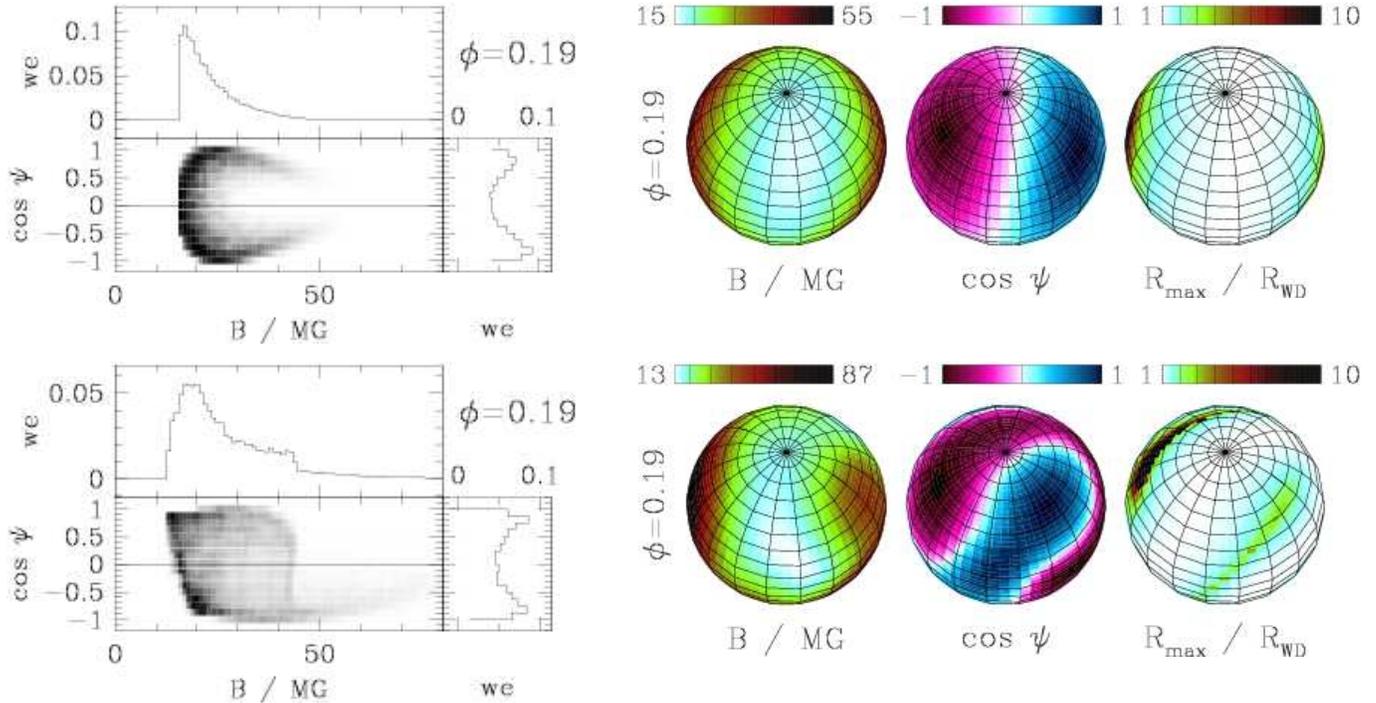

\includegraphics[width=7.6cm,clip]{6332.f4a}
\hfill
\raisebox{6mm}{\includegraphics[width=9.6cm,clip]{6332.f4b}}\\[1.5ex]
\includegraphics[width=7.6cm,clip]{6332.f4c}
\hfill
\raisebox{6mm}{\includegraphics[width=9.6cm,clip]{6332.f4d}}
\caption{Visualization of the two magnetic field models discussed for
\blhyi, the offset dipole (top) and the $l_\mathrm{max}=3$ multipole expansion
(bottom), both shown for orbital phase $\phi=0.19$ with the main
accretion region on the lower hemisphere in front. The \bpsi~diagrams
are on the left and the model field distributions on the
right. Further details are as in Fig.~2.}
\label{fig:bl}
\end{figure*}

For all three objects, the dipole, even after allowing for an
off-center shift in the three spatial coordinates, does not provide a
good fit, whereas the multipole expansions fare decidedly better.
While it is possible that substantially more complicated hybrid models
might be successful, we are limited in the number of models that could
be tested by the slow convergence properties of our code (see
Sect.~5). We quote some of the parameters of the dipole models in the
text and list the coefficients \glm\ and \hlm\ of the best-fit
multipole expansions \citep{euchneretal06} in
Tab.~\ref{tab:multipole}. These coefficients are given in MG and,
although such models are difficult to visualize, the numbers allow
some insight into the field structure: the three $l=1$ coefficients
combine to define the dipole which is allowed to be inclined relative
to the multipole axis; the $l,m=2,0$ coefficient describes the
quadrupole aligned along the multipole axis and the following $m=0$
coefficients the octupole and higher multipole zonal components; the
$m\ne 0$ (tesseral) components are modulated in azimuth in addition to
zenith angle.

Figures~3 to 5 contain the representations of the field structure at
$\phi\simeq \phi_0$ for both the offset dipole and the multipole
expansions. In these figures, progressing phase corresponds to
anti-clockwise rotation and a motion of a feature on the globes from
left to right. The most probable field strength seen at the face when
the prospective accretion spot faces the observer and the range of
field strength over the whole star are listed in
Tab.~\ref{tab:brange}. We discuss the results in these figures and
tables with each object below.

\subsection{\oef}


The main accretion region in \eferi\ faces the observer near phase
$\phi=0.10$ and is located $\sim 30^\circ$ from the rotational axis
\citep{beuermannetal87,piirolaetal87b}. The field strength in the
accretion spot is low as indicated by the featureless optical
cyclotron continuum. Zeeman halo absorption in the ordinary ray of the
cyclotron continuum \citep{oestreicheretal90} and cyclotron humps in
the infrared \citep{ferrarioetal96,harrisonetal04,howelletal06}
suggest a field in the range of 10 to 21\,MG with different studies
favoring different values. The positive circular polarization in the
high state \citep{piirolaetal87b,oestreicheretal90} implies that the
accreting field line in the main (X-ray emitting) spot points away
from the observer (negative cos\,$\psi$). There is evidence from
spectropolarimetry \citep{oestreicheretal90} and from broad-band
polarimetry \citep{piirolaetal87b} that a second accretion region is
located near the same meridian at a colatitude of about $115^\circ$
(lower hemisphere) with a disputed polarity. While
\citet{piirolaetal87b} argue for the same polarity as the main spot,
\citet{oestreicheretal90}~ favor opposite polarity. Both accretion
regions are only about $80^\circ$ apart, an observation that led
\citet{meggittwickramasinghe89} to suggest a quadrupole field.

The best-fit offset dipole (model A in Fig.~\ref{fig:spec1} and top
panels in Fig.~\ref{fig:ef}) has a polar field strength (of the
unshifted dipole) of 44.0\,MG, is inclined to the rotation axis by
$75^\circ$, and is shifted off center by $x_\mathrm{d}$,
$y_\mathrm{d}$, and $z_\mathrm{d}$ equaling 0.26, --0.01, and
--0.17\,\rwd, respectively, with \rwd\ the white dwarf radius. The
$z$-offset is directed along the dipole axis, $x$ and $y$ denote
perpendicular directions \citep{euchneretal02}. A moderately
good fit at $\phi=0.14$ it accompanied by an utter failure at
$\phi=0.89$ with a global \chisqred=19.97. This is seen in the
\halpha\ and \hbeta\ $\sigma^-$ features in the flux spectrum of
Fig.~\ref{fig:spec1} at 4800 and 6100\AA\ and in the polarization
spectrum at all wavelengths. As an added disadvantage, the model lacks
open field lines at the position of the accretion spot near the
meridian in the upper hemisphere at $\phi=0.14$ (Fig.~\ref{fig:ef},
top panel). As expected, setting the shifts perpendicular to the axis
equal to zero does not improve the situation. A more complex hybrid
model of aligned dipole, quadrupole and octupole, centered or
off-centered, did not fare well either. Our fits quite reliably
exclude this class of models.

\begin{figure*}[t]
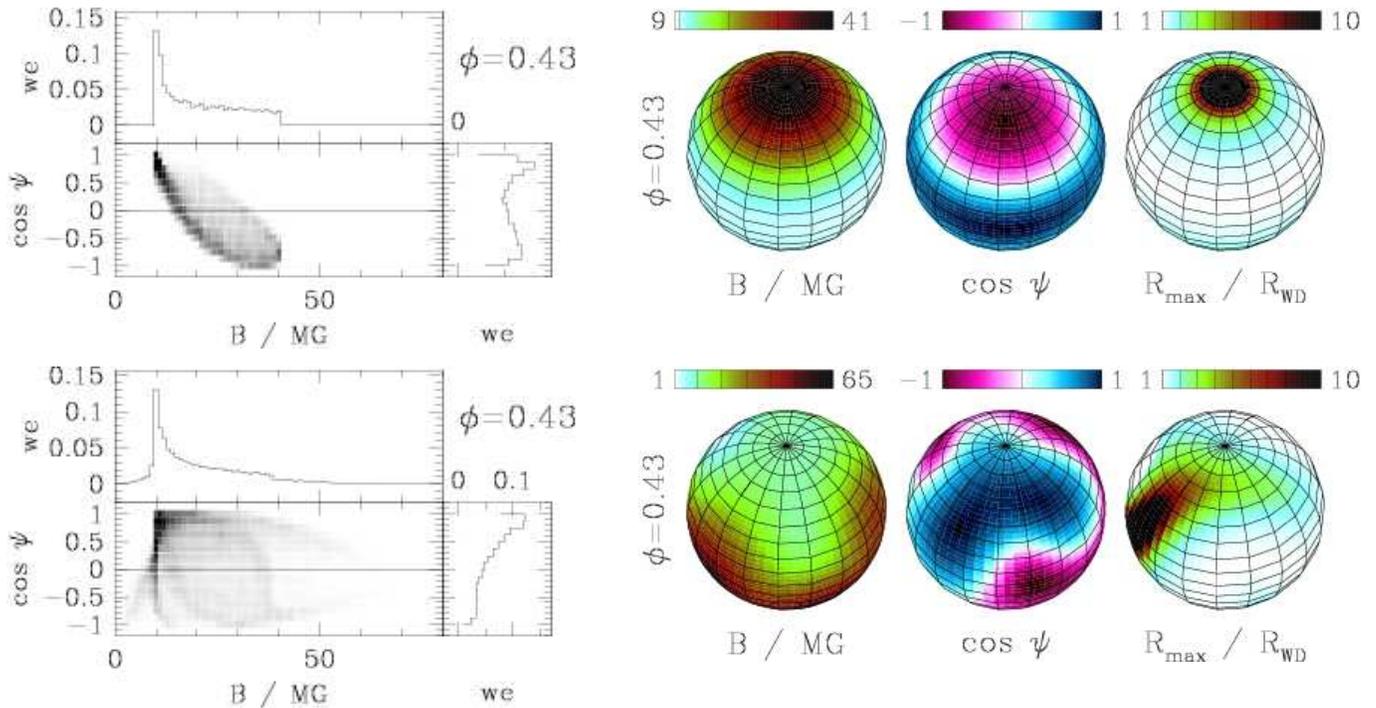

\includegraphics[width=7.6cm,clip]{6332.f5a}
\hfill
\raisebox{6mm}{\includegraphics[width=9.6cm,clip]{6332.f5b}}\\[1.5ex]
\includegraphics[width=7.6cm,clip]{6332.f5c}
\hfill
\raisebox{6mm}{\includegraphics[width=9.6cm,clip]{6332.f5d}}
\caption{Visualization of the two magnetic field models discussed for
\cptuc, the offset dipole (top) and the $l_\mathrm{max}=3$ multipole expansion
(bottom), both shown for orbital phase $\phi=0.43$ with the main
accretion region on the lower hemisphere in front. The \bpsi~diagrams
are on the left and the model field distributions on the
right. Further details are as in Fig.~2.}
\label{fig:cp}
\end{figure*}

Of the multipole expansions up to $l_\mathrm{max}=3$ (with 15 fit
parameters) and $l_\mathrm{max}=5$ (with 35 fit parameters), the
latter yields a slightly better fit with \chisqred=13.99~vs. 14.88 for
the former. The $l_\mathrm{max}=5$ case is shown in
Fig.~\ref{fig:spec1} as model B. There are some subtle differences
resulting in a better fit of either model to one or the other spectral
feature and, in summary, one would have to conclude that there is no
good basis for including the added parameters of the
$l_\mathrm{max}=5$ model if judgment is based solely on the flux and
polarization spectra. In line with this, we find that the
\bpsi~diagrams in Fig.~\ref{fig:ef} (center panel for
$l_\mathrm{max}=3$; bottom panel for $l_\mathrm{max}=5$) are similar
in the predominance of field strengths of 10--15\,MG, but differ in
the extensions to higher field strengths. The decisive difference of
the two multipole expansions is depicted in the rightmost globes in
Fig.~\ref{fig:ef}.  The $l_\mathrm{max}=3$ model possesses only a
small spot from which field lines reach far out. Such field lines,
whether they close within the Roche lobe of the white dwarf or not are
needed for the accretion stream to be guided towards the white
dwarf. The field in this tiny spot is directed outward (cos\,$\psi
>0$), however, leading to negative polarization of the cyclotron
emission from this potential accretion spot, while the observed
polarization is positive (cos\,$\psi <0$). The $l_\mathrm{max}=5$
model, for comparison, displays a large arc-like region of ingoing
field lines which covers also the expected location of the primary
accretion region $\sim 30^\circ$ from the rotation axis and facing the
observer shortly past $\phi=0$
\citep{beuermannetal87,piirolaetal87b}. This ribbon of open ingoing
field lines with cos\,$\psi <0$ winds around a region of
quadrupole-like low-lying magnetic arcs in the lower right quadrant at
$\phi=0.14$\,\footnote{The interested reader may look up Fig.~7 of
\citet{schwopeetal95} which shows the field structure of such a region
at its upper pole.}.  The decisive point in favor of the
$l_\mathrm{max}=5$ truncated multipole expansion among the models
studied is the correct polarity of the open field lines with
cos\,$\psi<0$ and a positive sign of the resulting circular
polarization of the high-state cyclotron emission from the main
accretion spot.

As seen from Tab.~\ref{tab:multipole}, the dipole component of the
multipole expansion is relatively weak with a polar field strength of
8.5\,MG obtained by squaring the three dipole coefficients. The
strongest components are azimuthally modulated quadrupole-like
ones. All individual $l=3-5$ coefficients are smaller than 10\,MG, but
their combined effect is significant in shaping the field structure
(Fig.~\ref{fig:ef}) which is not that of an $m=0$ quadrupole. Hence,
from the present study, we conclude that the field structure of \oef\
is substantially more complex than that of a centered or offset dipole
or quadrupole and may be even more complex than suggested by the
present best fit.

\subsection{\obl}

\blhyi\ is another polar that displays a complex accretion geometry,
usually referred to as 'one-pole' and 'two-pole' accretion in states
of low and high accretion rates, respectively
\citep{beuermannschwope89}. The main hard X-ray and cyclotron emitting
accretion spot is located in the lower hemisphere of the white dwarf
and is visible only for part of the orbit. Its appearance at the limb
of the white dwarf defines photometric phase $\phi=0$. It slowly
disappears some 0.40 phase units later \citep{piirolaetal87a}. The
second emission region in the upper hemisphere facing the observer
emits in the soft X-ray and EUV regime and is visible over much of the
orbital period \citep{schwopebeuermann93,szkodyetal97}. The sign of
the circular polarization of the cyclotron emission from the main
(second) pole is negative (positive) implying positive (negative)
cos\,$\psi$ \citep{cropper87,bailey88,schwopebeuermann89}. The
dominant photospheric field strength is 22\,MG, but the presence of
significantly higher fields is inferred from Zeeman spectroscopy. A
field of only 12\,MG was at times detected by $\halpha$ Zeeman
absorption in the cyclotron continuum emission of the main accretion
spot \citep{schwopeetal95}, but the feature may have originated at
some height above the white dwarf surface.

Our present data cover the whole binary orbit in five about equally
spaced intervals. The best-fit offset dipole model features a polar
field strength of 59.5\,MG, an inclination against the rotation axis
of $80^\circ$, and offsets of --0.18, 0.15, and --0.03 \rwd\ in $x,
y$, and $z$.  As in the case of \eferi, the offset dipole model does
not fare well. We find \chisqred=200.3 which is so large because of
the small standard deviation in the low-noise observed spectrum and
obviously poor fits in a few places. The model fails to reproduce the
\halpha\ $\sigma^-$ feature in the flux spectra near 6000\AA\ and also
much of the detail in the circular polarization spectra over the
entire wavelength range (Fig.~\ref{fig:spec2}, left panel). The fit to
the polarization data is particularly poor at $\phi=0.40$ (not
shown). We discard the offset dipole model because it provides a 
comparatively poor fit and because of the lack of open field lines
in both the upper and the lower hemisphere at $\phi=0.19$ at positions
which might correspond to the main and the secondary accretion regions
(Fig.~\ref{fig:bl}, upper right).

The multipole expansion truncated at $l_\mathrm{max}=3$ fits better
than the offset dipole, although there are still some systematic
deviation in the flux and circular polarization spectra, in particular
near 4300\AA\ as well as 6050\AA\ and around 5500\AA, respectively.
In spite of the still very high \chisqred=193.51, the field
distribution looks promising. On the positive side, we note the
correct representation of the reversal of the polarization near
4600\AA\ between $\phi=0.19$ and $\phi=0.81$ and the approximately
correct description of the \halpha\ $\sigma^-$ polarization near
6100\AA. As in case of \eferi, the quadrupole-like components are
strongest but, again, the other components are essential for the field
structure (Tab.~\ref{tab:multipole}). There is a longish ribbon of
field lines reaching out moderately far (about 3\,\rwd). It faces the
observer near $\phi=0.19$ and yields negative circularly polarized
cyclotron emission (positive cos\,$\psi$) as required for the main
accretion region. A second region with negative cos\,$\psi$ may allow
access as close as 35$^\circ$ from the rotational pole and may be
responsible for the intense flaring soft X-ray emission
\citep{schwopebeuermann89} and the positive circularly polarized
cyclotron emission at orbital phases when the main spot is behind the
white dwarf. The \bpsi~diagram prominently shows field strengths
between 13 and 45\,MG with a faint (and ill-defined) extension to
beyond 100\,MG. The ribbon which may contain the main accretion spot
crosses the 13\,MG field minimum quite consistent with the featureless
cyclotron continuum and the 12-MG Zeeman absorption in the cyclotron
emission of the main spot. These facts suggest that the model
approaches reality, although the large \chisqred\ lets us suspect that
the final model is still different. The experience from \eferi\ with
substantially different structures of the best fit multipole
expansions for $l_\mathrm{max}=3$ and $l_\mathrm{max}=5$ suggests that
such an improved model can be found. In this case, however,
convergence problems have prevented the construction of an
$l_\mathrm{max}=4$ or 5 model. We do not expect such a better-fitting
model to possess a simpler structure than the $l_\mathrm{max}=3$ model
and conclude that \blhyi\ possesses a complex field geometry, probably
not completely unraveled with the model presented here.

\subsection{\ocp}

Unlike the majority of polars, \ocp\ was discovered by its hard X-ray
emission \citep{misakietal96}. It displays a broad dip in the X-ray
flux which was used to derive a rotational ephemeris of the white
dwarf \citep{ramsayetal99}. The narrow emission line from the heated
face of the secondary star yields the orbital period
\citep{thomasreinsch96} which agrees with the rotational period
proving synchronism.  The ephemeris of these authors shows that
inferior conjunction of the secondary star occurs at dip phase
$\phi=0.08$. There is debate about the location of the accretion
region: \citet{misakietal96} suggested that the energy-dependent X-ray
dip arises from photoabsorption; while \citet{ramsayetal99} assign it
to a self-eclipse by the white dwarf. In the first case, the accretion
spot conveniently faces the observer shortly before inferior
conjunction; in the latter it faces away from the secondary.

As noted above, our data cover only one half of the binary orbit. They
were combined into two sets of flux and polarization spectra at dip
phases $\phi=0.99$ and 0.21 plus a noisier single set at
$\phi=0.43$. In the Misaki et al. and Ramsay et al. interpretations,
the accretion spot faces the observer near $\phi=0$ and $\phi\simeq
0.55$, respectively. We show our spectra at $\phi=0.99$ and $\phi=
0.43$ in Fig.~\ref{fig:spec2} (right panel). Model A is the offset
dipole and model B the multipole expansion with $l_\mathrm{max}=3$ of
which the latter yields a formally better fit with \chisqred
\,=\,12.35 and 11.78, respectively. Visual inspection, however, shows
that the differences are not pronounced and that both models reproduce
the major Zeeman features of \hbeta\ and the higher Balmer lines in
the flux and the polarization spectra reasonably well. The \bpsi\,
diagrams in Fig.~\ref{fig:cp} demonstrate the preponderance of field
strengths close to 10\,MG which make \cptuc\ a low-field polar and may
explain the hard X-ray spectrum. The offset dipole has a polar field
strength of 19.8\,MG, is practically aligned with the rotation axis,
and offset mainly in $z$ by 0.21\,\rwd. It has its high-field pole in
the upper hemisphere, whereas the multipole model possesses low fields
in the same region. The dipole component of the multipole expansion
has a polar field strength of 17.8\,MG, but the quadrupole-like
components are of similar strength and the octupole-like components
are non-negligible (Tab.~\ref{tab:multipole}). Both models, offset
dipole and multipole expansion, differ in the sign of the longitudinal
field component over the near and far hemisphere. The featureless
cyclotron continuum indicates a low field strength
\citep{thomasreinsch96} and the negative circular polarization of the
continuum \citep{ramsayetal99} shows that the accretion region must lie
in the lower (upper) hemisphere for the dipole (multipole) model. Only
the multipole model, however, possesses an extended region of outgoing
field lines which faces the observer near $\phi=0.55$ as expected from
the Ramsay et al. model (some 0.12 in phase later than in the lower
right globe of Fig.~\ref{fig:cp}). In summary, \cptuc\ is the third of
the polars studied here with a field structure more complex than a
simple offset dipole. As a caveat we recall the incomplete phase
coverage. The results for \cptuc\ are, therefore, preliminary.

\subsubsection{Summary of results}

For all three objects, the truncated multipole expansions yield
significantly better fits than the offset dipole models. Only the
former provide access to the surface of the white dwarf via field
lines reaching sufficiently far out at the expected positions on the
surface. As seen from Tab.~\ref{tab:brange}, dipole and multipole
models have practically the same most frequent values of the field
strength. They cover also similar ranges, in particular, if one
considers that the faint extensions in the probabiltiy distribution to
the lowest and highest field strengths are not well defined. This
similarity reflects the fact that both models more or less fit the
principal features of the Zeeman spectra. Intuitively, one might
consider rotating the offset dipole model to match the accretion
conditions, but that changes the phase-dependent \bpsi~diagrams and
destroys the fits to the Zeeman spectra. Hence, the offset dipoles
clearly cannot match all conditions simultaneously. In spite of the
similar $B$-distributions, the field structures of the two models
\emph{are} significantly different and our conclusion in favor of
stuctures more complex than offset dipoles is safe.

\section{Discussion}

We have presented the first Zeeman tomographic study of the field
structure of white dwarfs in polars based on phase-resolved VLT
spectropolarimetry. We have demonstrated that the studied stars
possess field structures significantly more complex than simple
centered or offset dipoles. Such a result was considered possible or
even likely in many previous publications, but detailed proof was not
so far available. Our results clearly demonstrate this complexity,
although we have to caution that our best fits may not yet describe
reality in every detail. The simplest parameter indicative of a
structure more complex than a centered dipole is the range of field
strengths over the surface of the star which exceeds seven for the
best fit models for all three stars, while it would be two for a
centered dipole. The presence of strong higher multipole components
besides the dipole may be surprising considering the fact that the
white dwarfs in CVs have a typical age in excess of 1\,Gyr
\citep{kolbbaraffe99} sufficient to expect substantial decay of the
higher order components \citep[e.g.][]{cumming02}. Recreation of
higher order poloidal components from a toroidal interior field has
been suggested by \citet{muslimovetal95}, but it is not known whether
the required strong toroidal field exists in magnetic white
dwarfs. The single white dwarfs HE\,1045-0908 and PG\,1015+014
\citep{euchneretal05,euchneretal06} have similarly complex magnetic
field structures at an age of $\sim 0.5$\,Gyr. At present, field
evolution is not sufficiently constrained by observations, but may
become so when when the field structure of more objects becomes
available.

We find that fitting the field structure of accreting white dwarfs in
CVs offers a decisive advantage over the analysis of isolated white
dwarfs. The location of the accretion spot and the absolute value and
direction of the field vector in the spot can be deduced
independently, e.g., from X-ray and optical light curves and from
broad-band polarimetry. Requiring that a successful field model
complies with this independent information turns out to be a powerful
tool and is the main driver for our conclusion in favor of field
structure more complex than an offset dipole in all three
objects.

There is a semantic aspect worth mentioning. Comparison of the
right-hand globes for the multipole expansions and the offset dipoles
in Figs.~3--6 (including the phases omitted for conciseness)
demonstrates that the latter have well-defined circular regions of
outgoing field lines which are properly addressed as `poles'. The
corresponding regions in the multipole structures, on the other hand,
are quite irregular and longish structures which imply accretion
geometries which probably allow access to the white dwarf surface at
more than one position characterized by a wide range of angular
separations. The resulting accretion geometry is no longer
appropriately described by the dipole-inspired expressions `one-pole
accretion' or `two-pole accretion'.

Given the complicated field structures in the three polars studied
here, it is desirable to extend the Zeeman tomographic analysis to a
larger number of objects in order to distinguish between
idiosyncrasies of individuals and the general properties of the
class. Such a program is feasible, but it calls also for a
consideration of the limitations of our approach. One obvious
limiting factor is telescope time. Although we have used typically two
orbital periods of high signal-to-noise spectra using the VLT and the
spectropolarimetric capabilities of FORS1, the remaining noise in the
circular polarization spectra limits the discrimination between the
\bpsi\ diagrams of different models. A more extensive tomographic
program requires to cover several orbital periods of each target with
an 8-m class telescope. A second limiting factor is CPU time. In order
to thoroughly test a given field model, typically 50 to 100
\chisq-minimization runs are required becauce they tend to
get stuck in secondary minima of the complicated \chisq\
landscape. Each step in this process requires to assemble the Zeeman
spectra for the respective field model from the database. The
resulting lack of speed is the main reason why we had to limit the
number of models tested.

In the previous papers of this series
\citep{euchneretal05,euchneretal06}, we have already discussed the
alternative approach of \citet{donatietal94}, the Zeeman Broadening
Analysis (ZEBRA). This method determines the best-fitting
\bpsi~diagrams for each orbital phase interval directly from the data
employing the Maximum Entropy Method (MEM) as a regularization
procedure. The advantage is predictably speed, the disadvantage is the
uncertainty whether the individual \bpsi~diagrams are compatible with
any global physical field model. The method faithfully reproduces
the distribution in the absolute value of $B$ over the visible
hemisphere at a given orbital phase, but substantially smears the
angle relative to the line of sight \citep{donatietal94}. If
interpreted in terms of a multipole model, the reconstructed
\bpsi-diagram then leads to spurious higher multipole
components. Furthermore, the requirement that the so derived diagrams
contain the \bpsi\ combination describing the accretion spot is easily
implemented, but the exact location of the spot on the star is not,
because the method forgoes imaging. On the other hand, testing a larger number of field
models for compatibility with the \bpsi~diagrams for the individual
phase intervals will probably be less time consuming than our
present approach, because the large database of Zeeman spectra need no
longer be accessed after these diagrams have been established. We
plan to further study the different approaches in order to find the
ultimately preferable one. 

\begin{acknowledgements}
This work was supported in part by BMBF/DLR grant \mbox{50\,OR\,9903\,6}.  
BTG was supported by a PPARC Advanced Fellowship.
\end{acknowledgements}

\bibliographystyle{aa}

\end{document}